\documentclass[12pt]{article}
\usepackage{a4wide}
\usepackage{amssymb}
\usepackage{graphicx}
\usepackage{cite}
\begin{document}
\Large
\begin{center}
\textbf{Modelling financial markets by the multiplicative sequence of trades}\\[1.25\baselineskip]
\small \textbf {V. Gontis\footnote{Corresponding author. Tel.
+370-5-2623725; fax +370-5-2125361.
\\
\textit{E-mail address}: gontis@ktl.mii.lt (V. Gontis)}
and B. Kaulakys}

\vskip 2mm Institute of Theoretical Physics and Astronomy,
Vilnius University \\ A.Go\v{s}tauto 12, LT-2600 Vilnius, Lithuania
\end{center}

\abstract{We introduce the stochastic multiplicative point process
modelling trading activity of financial markets. Such a model
system exhibits power-law spectral density $S(f) \propto 1/f^\beta$ ,
scaled as power of frequency for various values of $\beta$ between
0.5 and 2. Furthermore, we analyze the relation between the
power-law autocorrelations and the origin of the power-law
probability distribution of the trading activity. The model
reproduces the spectral properties of trading activity and
explains the mechanism of power-law distribution in real markets.
} \vskip 2mm \textbf{Keywords:} financial markets, stochastic modelling,
point processes, 1/f noise. 

\baselineskip=1\normalbaselineskip

\section{Introduction}

\label{intro} Power-laws are intrinsic features of the economic
and financial data. There are numerous studies of power-law
probability distributions in various economic systems
\cite{1,2,3,4,41,42,8,5}. The key result in recent findings is
that the cumulative distributions of returns and trading activity
can be well described by a power-law asymptotic behavior,
characterized by an exponent $\lambda \approx 3$, well outside the
Levy stable regime $0<\lambda <2$ \cite {5}.

The time-correlations in the financial time series are studied
extensively as well \cite{5,6,7}. Gopikrishnan \emph{et al} \cite
{5,6} provided empirical evidence that the long-range correlations
for volatility were due to the trading activity, measured by a
number of transactions $N$.

Recently we adapted the model of $1/f$ noise based on the Brownian
motion of time interval between subsequent pulses, proposed in
\cite{9,10,11,12}, to model the share volume traded in the
financial markets \cite{13}. The idea to transfer long time
correlations into the stochastic process of the time interval
between trades or time series of trading activity is in
consistence with the detailed studies of the empirical financial
data \cite{5,6} and fruitfully reproduces the spectral properties
of the financial time series \cite{13,131}. Further, we
generalized the model defining stochastic multiplicative point
process to reproduce a variety of self-affine time series
exhibiting the power spectral density $S(f)$ scaling as a power of
frequency, $S(f) \propto f^{-\beta}$ \cite{132}.

In this contribution we analyze the applicability of the
stochastic multiplicative point process as a model of trading
activity in the financial markets. We investigate the spectral density
and counting statistics of model trading activity in comparison
with empirical data from the stock exchange. The model reproduces the
spectral properties of trading activity and explains the mechanism
of power law distribution in the real markets.

\section{The model}

We consider a point process $I(t)$ as a sequence of the $\delta$-type
random correlated pulses,
\begin{equation}
\label{eq:1}I(t)=\sum\limits_ka_k\delta (t-t_k),
\end{equation}
and define the number of trades $N_j$ in the time intervals
$\tau_d$ as an integral of the signal, $N_j=\int\limits_{t_j}^{t_j+\tau
_d}I(t)dt$. Here $a_k$ is a contribution of one transaction.
When $a_k=1$, the signal (\ref{eq:1}) counts the
transactions in the financial market. When $a_k$ describes asset price
change during one transaction, the signal counts the price changes.
When $a_k=\bar a$ is a constant, the point process is
completely described by the set of times of the events $\{t_k\}$
or equivalently by the set of interevent intervals $\{\tau
_k=t_{k+1}-t_k\}$. Various stochastic models for $\tau _k$ can be
introduced to define a stochastic point process. In papers
\cite{9,10,11,12} it has been shown analytically that the
relatively slow Brownian fluctuations of the interevent time $\tau
_k$ yield $1/f$ fluctuations of the signal (\ref{eq:1}). In
\cite{132} we have generalized the model introducing stochastic
multiplicative process for the interevent time $\tau _k$,
\begin{equation}
\label{eq:5}\tau _{k+1}=\tau _k+\gamma \tau _k^{2\mu -1}+\tau
_k^\mu \sigma \varepsilon _k.
\end{equation}
Here the interevent time $\tau _k$ fluctuates due to the external
random perturbation by a sequence of uncorrelated normally
distributed random variable $\{\varepsilon _k\}$ with zero
expectation and unit variance, $\sigma $ denotes the standard
deviation of the white noise and $\gamma \ll 1$ is a damping
constant. From the big variety of possible stochastic processes we
have chosen the multiplicative one, which yields multifractal
intermittency and power-law probability distribution functions.
Pure multiplicativity corresponds to $\mu=1$. Other values
of $\mu$ may be considered, as well.

The iterative relation (\ref {eq:5}) can be rewritten as Langevine
stochastic differential equation in $k$-space
\begin{equation}
\label{eq:6}\frac{d\tau _k}{dk}=\gamma \tau _k^{2\mu -1}+\tau
_k^\mu \sigma \xi \left( k\right).
\end{equation}
Here we interpret $k$ as continuous variable while $\left\langle
\xi \left( k\right) \xi \left( k^{\prime }\right) \right\rangle
=\delta (k-k^{\prime })$.

The steady state solution of the  stationary Fokker-Planck
equation with zero flow, corresponding to (\ref {eq:6}), gives the
probability density function for $\tau _k$ in the $k$-space (see,
e.g., \cite{19})
\begin{equation}
\label{eq:7}P_k(\tau
_k)=C\tau_k^\alpha=\frac{\alpha+1}{\tau_{max}^{(\alpha+1)}-\tau_{min}^{(\alpha+1)}}\tau
_k^\alpha ,\quad \alpha =2\gamma /\sigma ^2-2\mu.
\end{equation}

The steady state solution (\ref{eq:7}) assumes Ito convention
involved in the relation between expressions (\ref{eq:5}),
(\ref{eq:6}) and (\ref{eq:7}) and the restriction for the diffusion
$0<\tau_{\min }<\tau_k<\tau _{\max }$.

We have already derived  the formula for the power spectral density of the
multiplicative stochastic point process model, defined by
Eqs.~(\ref {eq:5}) and (\ref{eq:6}) for the interevent time
\cite{132},
\begin{equation}
S_{\mu }(f)=\frac{2C\overline{a}^{2}}{\sqrt{\pi }\overline{\tau
}(3-2\mu )f}\left(\frac{\gamma }{\pi f}\right)^{\frac{\alpha
}{3-2\mu }}{\mathop{\mathrm{Re}}}\int\limits_{x_{\min }}^{x_{\max
}}\exp \left\{-i\bigg(x-\frac \pi
4\bigg)\right\}{\mathop{\mathrm{erfc}}} (\sqrt{-ix})x^{\frac\alpha
{3-2\mu }-\frac 12}dx \label{eq:12}
\end{equation}
where $\bar \tau =\left\langle \tau _k\right\rangle =T/(k_{\max
}-k_{\min })$ is the expectation of $\tau _k$. Here we introduce
the scaled variable $x=\pi f \tau ^{3-2\mu }/\gamma$ and
$x_{\min }=\pi f \tau _{\min }^{3-2\mu }/\gamma,\quad x_{\max }=
\pi f \tau _{\max }^{3-2\mu }/\gamma$.

Expression (\ref{eq:12}) is appropriate for the numerical calculations
of the power spectral density of the generalized
multiplicative point process defined by Eqs. (\ref{eq:1}) and
(\ref{eq:5}). In the limit $\tau _{\min }\rightarrow 0$ and $\tau
_{\max }\rightarrow \infty $ we obtain an explicit expression
\begin{equation}
\label{eq:14}S_\mu (f)=\frac{C\overline{a}^2}{\sqrt{\pi
}\overline{\tau } (3-2\mu )f}\left(\frac \gamma {\pi
f}\right)^{\frac \alpha {3-2\mu }}\frac{\Gamma (\frac 12+\frac
\alpha {3-2\mu })}{\cos (\frac{\pi \alpha }{2(3-2\mu )})}.
\end{equation}

Equation (\ref{eq:14}) reveals that the multiplicative point
process (\ref{eq:5}) results in the power spectral density
$S(f)\sim 1/{f^\beta }$ with the scaling exponent
\begin{equation}
\label{eq:15}\beta =1+\frac{2\gamma /\sigma ^2-2\mu }{3-2\mu }.
\end{equation}

\begin{figure}[tbp]
\begin{center}
\includegraphics[width=0.7\textwidth]{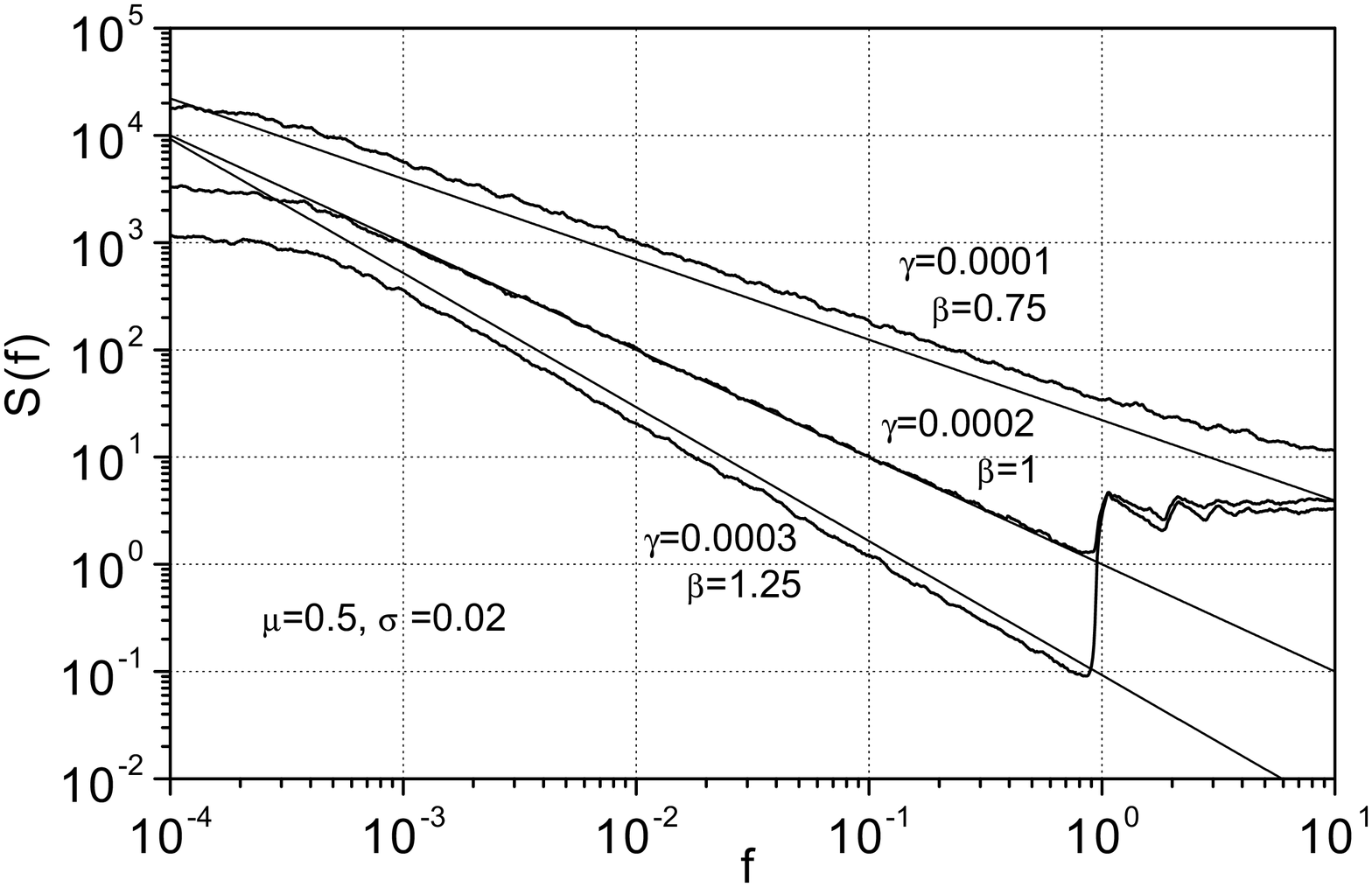}
\end{center}
\caption{Power spectral density $S(f)$ vs frequency $f$ calculated
numerically according to Eqs. (\ref{eq:1}) and (\ref{eq:5}) with the parameters
$\mu=0.5$, $\sigma=0.02$ and different relaxations of the signal
$\gamma$. We restrict the diffusion of the interevent time in the
interval $\tau _{min}=10^{-6}$; $\tau _{max}=1$ with the
reflective boundary condition at $\tau _{min}$ and transition to
the white noise, $\tau _{k+1}=\tau _{max}+\sigma\varepsilon_k$,
for $\tau_k>\tau _{max}$. The straight lines represent the results given
by the explicit formula (\ref{eq:14}).}
\label{fig:1}
\end{figure}

Let us compare our analytical results (\ref{eq:12}) and
(\ref{eq:14}) with the numerical calculations of the power
spectral density according to equations (\ref{eq:1}) and
(\ref{eq:5}). In Fig.~\ref{fig:1} we present the numerically
calculated power spectral density $S(f)$ of the signal $I(t)$ for
$\mu =0.5$ and $\alpha =2\gamma/\sigma ^2-1=0$, -0.5 and +0.5.
Numerical results confirm that the multiplicative point process
exhibits the power spectral density scaled as $S(f)\sim 1/f^\beta
$. Equation (\ref{eq:12}) describes the model power spectral
density very well in a wide range of parameters. The explicit
formula (\ref{eq:14}) gives a good approximation of power spectral
density for the parameters when $\beta \simeq 1$.  These results
confirm the earlier finding \cite {9,10,11,12} that the power
spectral density is related to the probability distribution of the
interevent time $\tau _k$ and $1/f$ noise occurs when this
distribution is flat, i.e., when $\alpha =0$.

It is likely that such a stochastic model with parameters in the
region $0.5\leq \beta \leq 1.5$ may be adaptable for a wide
variety of different systems. In this paper we will discuss
applicability of the model for the financial market.

We derived pdf of $N$ for the pure multiplicative model with $\mu
=1$ in \cite{132}

\begin{equation}
P(N)=\frac{C'\tau_{d}^{2+\alpha}(1+\gamma N)}{N^{3+\alpha}
(1+\frac{\gamma}{2}N)^{3+\alpha}}\sim\left\{
\begin{array}{ll}
\frac{1}{N^{3+\alpha}},& N\ll \gamma^{-1}, \\
\frac{1}{N^{5+2\alpha}},& N\gg \gamma^{-1}.
\end{array}
\right.
\label{eq:18}
\end{equation}

Probability distribution function for $N$ obtained from the
numerical simulation of the model is in a good agreement with the
analytical result (\ref{eq:18}).

\section{Discussion and conclusions}

We have introduced a multiplicative stochastic model for the time
intervals between events of point process. Such a model of time
series has only a few parameters defining the statistical
properties of the system, i.e., the power-law behavior of the
distribution function and the scaled power spectral density of the
signal. The ability of the model to simulate $1/f$ noise as well
as to reproduce signals with the values of power spectral density
slope $\beta$ between 0.5 and 1.5 promises a wide variety of
applications of the model.

\begin{figure}[tbp]
\begin{center}
\includegraphics[width=0.3\textwidth]{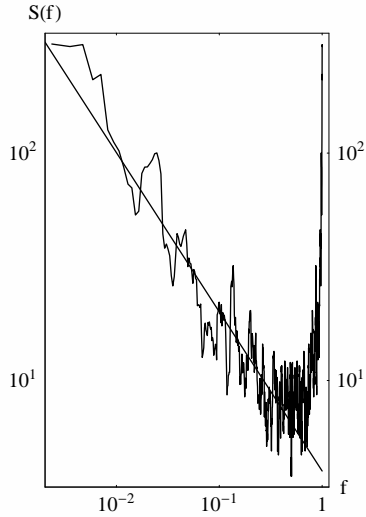}
\end{center}
\caption{Power spectral density $S(f)$ vs frequency $f$ calculated
numerically from the empirical data of 3 most liquid stocks from the
Lithuanian Stock Exchange. Straight line fits $S(f)\propto f^{-0.7}$.}
\label{fig:2}
\end{figure}

Let us present shortly the possible interpretations of the
empirical data of the trading activity in the financial markets.
With a very natural assumption of transactions in the financial
markets as point events we can model the number of transactions
$N_j$ in equal time intervals $\tau _d$ as the outcome of the
described multiplicative point process. We already know from
available studies \cite{5} that the empirical data exhibit power
spectral density in the low frequency limit with the slope $\beta
\simeq 0.7$. Empirical data from the Lithuanian Stock Exchange for
the most liquid assets confirm the same value $\beta \simeq 0.7$,
(see Fig~\ref{fig:2}). For the pure multiplicative model with
$\mu =1$ this results in $\alpha =2\gamma /\sigma ^2-2\mu
\simeq -0.3$. The corresponding cumulative distribution of $N$ in
the tail of high values (see equation (\ref{eq:18})) has the
exponent $\lambda=4+2\alpha =3.4$. This is in an excellent
agreement with the empirical cumulative distribution exponent 3.4
defined in \cite{5} for 1000 stocks of the three major US stock
markets.

The numerical results confirm that the multiplicative stochastic
model of the time interval between trades in the financial market
is able to reproduce the main statistical properties of trading
activity $N$ and its power spectral density. The power-law
exponents of the pdf of the interevent time, $\alpha $, and the
cumulative distribution of the trading activity, $\lambda $, as
well as the slope of power spectral density, $\beta$, are defined
just by one parameter of the model $2\gamma /\sigma ^2$. The model
suggests a simple mechanism of the power-law statistics of trading
activity in the financial markets.
%

\end{document}